\documentclass[10pt]{article}
\usepackage{multicol,ftnright,citesort,epsfig}
\begin{document}

\title{Adsorption of polydisperse polymer chains}

\author{{\bf Richard P. Sear}\\
~\\
Department of Physics, University of Surrey\\
Guildford, Surrey GU2 5XH, United Kingdom\\
email: r.sear@surrey.ac.uk}

%\date{\today}

\maketitle

\begin{abstract}
The adsorption of polydisperse ideal polymer chains is shown to be
sensitive to the large $N$ tail of the distribution of chains. If and only if
the number of chains decays more slowly than exponentially then there
is an adsorption transition like that of monodisperse infinite chains.
If the number decays exponentially the adsorption density diverges
continuously at a temperature which is a function of the mean chain length.
At low coverages, chains with repulsive monomer--monomer interactions
show the same qualitative behaviour.
\end{abstract}

%36.20.Cw Molecular weights, dispersity
%36.20.-r Macromolecules and polymer molecules
%61.25.Hq Macromolecular and polymer solutions; polymer melts; swelling
%67.70.+n Films (including physical adsorption)
\vspace*{0.1in}

\begin{multicols}{2}
%\newpage

Synthetic polymers are almost inevitably polydisperse. The polymer
chains are not all of the same length, and so a polymer is
not a single component but a
mixture of chains of different lengths.
However, almost all of the large
number of theoretical
studies of polymer adsorption have considered monodisperse
polymer chains, where the chains are all of the same length. This is done
to simplify matters and is a reasonable approximation when, as is often
the case, the width of the distribution
of lengths of polymer chain is small. Here we
study polymers in which there are polymer chains with a very
wide range of lengths, paying particular attention to the longest
polymers in the distribution.
Our motivation is not just the experimental fact that synthetic polymers
are polydisperse yet are almost invariably treated within theory
as being monodisperse. Polydisperse polymers may, and we show that they do,
exhibit behaviour which is qualitatively different from that of
monodisperse polymers.
We study a very highly idealised model of polymer adsorption:
ideal chains adsorbing onto a wall due to a short-ranged attraction between the
wall and the monomers.
This problem has been extensively studied for monodisperse chains
\cite{rubin65,eisenriegler82,eisenriegler83,degennes,eisenriegler}, but not
as far as the author is aware for polydisperse chains.

The adsorption of an ideal chain is a textbook problem. It is highly
idealised: almost invariably the density of polymer adsorbed onto a
surface is too high for interactions between the polymer segments to be
neglected. However, it has served well as a simple first model of adsorption
of monodisperse polymer chains \cite{degennes}
and we shall use it as such for polydisperse chains.
There has been previous work within
the Scheutjens--Fleer theory \cite{roefs94},
on the adsorption of polydisperse polymer
chains but not, as far as the author is aware, on ideal chains.
Roefs {\it et al.} \cite{roefs94} did not consider the large $N$
tail of the distribution and so did not find the behaviour we will
describe below.
Note that in ref. \cite{roefs94}
the ratio between the volume of the solution and the
surface area of the wall was only of the order of the radius of
gyration of the polymers. By contrast here we study a wall in contact
with a bulk polymer solution so the ratio between the volume and
surface area is infinity.

We start by briefly reviewing the behaviour of monodisperse chains,
with particular emphasis on the dependence of the adsorption
on the length of the chain.
The chains are ideal and each consists of a linear chain of $N$
monomers of length $a$.
The chains are at a non-zero number density $\rho$ in the bulk.
The bulk polymer solution is in contact with a wall. This wall
attracts monomers via a short-ranged attraction; the range is
taken to be $a$ for simplicity. The strength of attraction is $\epsilon$.
Having discussed monodisperse chains we then generalise the theory
to describe an arbitrary polydisperse mixture of chains of differing lengths.
We show that not all polydisperse mixtures behave in the same
way: there are three qualitatively different behaviours possible.
Which behaviour a mixture exhibits depends on the large $N$ tail
of the distribution.
This is also true of the cloud-point curve of polydisperse polymers,
as was shown by \v{S}olc \cite{solc70}.
Recent work by the author on
the bulk phase behaviour of polydisperse hard spheres \cite{sear}
has found a similar sensitivity to the tail of the distribution.

Ideal chains do not interact with each other and so finite ideal chains
are independent systems with a finite number of degrees of freedom.
They therefore cannot exhibit a phase transition. For ideal chains
there is only an adsorption transition in the limit that the number
of monomers $N$ tends to infinity. We denote the transition temperature
of infinite chains by $T_a$. For finite
chains there is only a steep increase in the adsorbed density near $T_a$.
Most distribution functions for polydisperse mixtures tend to zero only
at infinity (e.g., the Schulz and distribution \cite{flory,salacuse82}),
which is to say that they contain an infinitesimal
density of infinitely large chains. The question then arises: as there
are infinitely long chains present
is there a phase transition?

Below the adsorption temperature $T_a$ large chains,
$N\rightarrow\infty$ are adsorbed onto the wall.
The chains
form a layer of height $D$ which can be easily estimated using a
scaling argument due to de Gennes \cite{degennes}.
We restrict ourself to the weak-adsorption regime \cite{degennes} where
$D\gg a$.
The free energy of adsorption of a single ideal chain has two competing
parts. The first is the entropy change when a polymer chain is confined
to a layer of height $D$, this scales as $-Na^2/D^2$. We take
Boltzmann's constant to be equal to unity. The second is the energy change due
to adsorption, this is $-N\epsilon$ times the fraction of monomers within
the range of wall's attraction $a/D$. So, the free energy of
adsorption $\Delta F$ is given by
\begin{equation}
\Delta F \sim T\frac{Na^2}{D^2}- N\epsilon \frac{a}{D}.
\label{df}
\end{equation}
The width $D$ of the layer at equilibrium is found by minimising
$\Delta F$
\begin{equation}
D\sim a(T/\epsilon),
\label{deq}
\end{equation}
which gives an adsorption free energy of
\begin{equation}
\Delta F\sim - N \epsilon^2/T.
\end{equation}
An exact treatment of adsorbed ideal chains yields
\cite{rubin65,eisenriegler82,eisenriegler83,eisenriegler}
\begin{equation}
\frac{\Delta F(N)}{T}\sim
- N\left(\epsilon/T-\epsilon/T_a\right)^2
~~~~~ T<T_a,\\
\label{fads}
\end{equation}
which is the dominant part of the free energy difference between
a chain in the bulk and a chain with one end near the wall,
when both $N$ and $|\Delta F|$ itself are much larger than unity
but we are still in the weak-adsorption regime so
$\epsilon/T-\epsilon/T_a$ is small.
For large chains, the free energy difference $\Delta F$ between a chain
adsorbed onto the wall and one in the bulk increases linearly with $N$.

Given $\Delta F$ we can follow the adsorption by the number density
of chains adsorbed at the wall.
Because the chains are ideal their density in the presence of
an external field is just the bulk density times a 
Boltzmann weight $\exp(-\Delta F/T)$.
The free energy $\Delta F$ given by eq. (\ref{fads}) is
the exact form for a
polymer chain which has one end at the wall. Thus the density
of polymer chains with at least one end near the wall is
\begin{eqnarray}
\rho_a&\sim &\rho \exp\left(-\Delta F/T\right)\\
\label{rhoa2}
&\sim &\rho \exp\left[N(
\epsilon/T-\epsilon/T_a)^2\right]
~~~~~ T<T_a.
\label{rhoamono}
\end{eqnarray}
We can think of other densities such as the number density of monomers
within $a$ of the wall or the number density within the whole
adsorbed layer, but both these densities behave in qualitatively the same
way (as a function of temperature) as $\rho_a$.
Above the adsorption transition, $T>T_a$, the density near the wall is
less than in the bulk, and tends to zero close
to the wall for infinitely long chains.
In the $N\rightarrow\infty$ limit then, there is an, infinite, jump
in $\rho_a$ at $T_a$: there is a thermodynamic transition at $T=T_a$.
For finite $N$ there is a sharp increase in $\rho_a$ near $T=T_a$
but no transition.

The generalisation to polydisperse ideal chains is easy. In the bulk
the density of polydisperse chains of different lengths $N$ is determined
by a distribution function $x(N)$. The number density of
chains of length $N$ is $\rho x(N){\rm d}N$ \cite{salacuse82},
where $x$ is normalised
\begin{equation}
\int_0^{\infty}x(N){\rm d}N=1
\end{equation}
so that the total number
density of chains is $\rho$ as before. Thus the density of monomers
of chains of length $N$ near the wall is
$\rho x(N)\exp[-\Delta F(N)/T]{\rm d}N$ and so the adsorbed
density $\rho_a$ when the polymer chains are polydisperse is
given by the obvious generalisation of eq. (\ref{rhoa2}),
\begin{equation}
\rho_a\sim\rho \int_0^{\infty}
x(N)\exp[-\Delta F(N)/T]{\rm d}N
~~~~~ T<T_a.
\label{rhoapolydef}
\end{equation}
This is a definition, to obtain the behaviour we insert the
free energy expression eq. (\ref{fads}) which is asymptotically exact
in the $N\rightarrow\infty$ limit, in the weak adsorption regime.
This is sufficient as it is the behaviour
in this limit that determines whether or not there is an thermodynamic
transition --- finite chains cannot contribute to a discontinuity
in $\rho_a$ or its derivatives. So, using eq. (\ref{fads}) in
eq. (\ref{rhoapolydef})
\begin{equation}
\rho_a\sim\rho \int_0^{\infty}
x(N)\exp\left[N(\epsilon/T-\epsilon/T_a)^2\right]{\rm d}N
~~~~~ T<T_a.
\label{rhoapolydef2}
\end{equation}

For monodisperse infinitely long chains then the adsorbed density is
infinite below $T_a$. We see that for polydisperse chains
this is also true if the integral of eq. (\ref{rhoapolydef2}) does not
converge, which it does not whenever $x(N)$ decays more slowly then
exponentially with $N$ {\it or} it does decay exponentially like
$\exp(-bN)$ ($b$ positive) but $b<(\epsilon/T-\epsilon/T_a)^2$.
So, if the decay of $x(N)$ is
slower than exponential, e.g., log-normal or power law, then
polydisperse chains behave as infinitely long chains below $T_a$.
There is therefore an adsorption transition at $T_a$.
If the decay of $x(N)$ is faster than exponential then
there are not enough of the infinitely long chains to cause a transition
and $\rho_a$ increases sharply but does not diverge.
An analogous dependence on $x(N)$ is found in the cloud-point
curve of a polymer solution \cite{solc70}. There if the decay
is slower than exponential, the cloud-point curve occurs at the
same effective temperature as that for infinite polymers. As here
a slower than exponential decay means that the density of
infinitely long polymers is high enough to cause the polydisperse solution
to behave as a solution of infinite monodisperse polymer.
The similarity in behaviour between the cloud-point curve and the
results presented here is due to the fact that the free energy
which determines the partitioning of the polymer at the cloud-point curve
is also linear in $N$.

But the most interesting case is that of an exponential decay. There
the behaviour is not like that of any monodisperse polymer.
Consider a simple exponentially decreasing
distribution function
\begin{equation}
x(N)={\overline N}^{-1}\exp(-N/{\overline N})
\label{xexp}
\end{equation}
where ${\overline N}$ is the average length of the polydisperse chains.
The exponential distribution
is a special case of the widely used Schulz distribution \cite{flory}.
Inserting the distribution function eq. (\ref{xexp}) into
eq. (\ref{rhoapolydef2})
for the adsorption we obtain
\begin{eqnarray}
\rho_a&\sim&
\frac{\rho }{{\overline N}}\int_0^{\infty}
\exp\left[N(\epsilon/T-\epsilon/T_a)^2-N/{\overline N} \right]{\rm d}N
\nonumber\\
\label{pgdef}\\&\sim&
\frac{\left(\rho /{\overline N}\right)}
{{\overline N}^{-1}-(\epsilon/T-\epsilon/T_a)^2}
~~~~~T_e<T<T_a.
\label{gamma}
\end{eqnarray}
Below $T_e$ $\rho_a$ is infinite.
\begin{equation}
T_e=
\frac{T_a}{1+{\overline N}^{-1/2}(T_a/\epsilon)}.
\label{ttran}
\end{equation}
For an exponentially distributed polydisperse mixture of
polymer chains the adsorbed density diverges continuously;
the density diverges unlike finite monodisperse chains and it
does so continuously, not via a jump as monodisperse infinite chains do.
The divergence
occurs at a temperature $T_e$ which is below the adsorption temperature
of infinite chains by an amount proportional to
${\overline N}^{-1/2}$.

The divergence is also present for a distribution function $x(N)$
which is an exponential times a power law, as the Schulz distribution 
function is \cite{flory,salacuse82}. Indeed the behaviour is
qualitatively unchanged if the exponential is multiplied by any
function of $N$ which varies more slowly than exponentially, allowing
the exponential dependence to dominate at large $N$. See ref.
\cite{flory} for different distribution functions found for linear
polymers.

All the above is for ideal polymers. What about polymers with monomers
which repel each other, the standard model of a polymer in a good solvent?
The free energy of adsorption of {\it isolated} polymer chains with monomers
which repel each other again has a dominant linear term in $N$
although the dependence on $T$ is different
\cite{eisenriegler82,eisenriegler}.
An exact field-theory treatment yields for long chains
with repulsive monomer--monomer interactions
\cite{eisenriegler82,eisenriegler}
\begin{equation}
\frac{\Delta F(N)}{T}\sim
- N\left(\epsilon/T-\epsilon/T_a\right)^{1/\Phi_0}
~~~~~ T<T_a,\\
\end{equation}
where $T_a$ is the adsorption temperature for interacting chains
and again this is the part of the adsorption free energy which dominates
when the free energy
of adsorption is negative and has a magnitude much larger than unity.
The exponent $\Phi_0$ is estimated to be 0.59 \cite{eisenriegler}.
So, $\Delta F$ is linear in $N$ as before
and so isolated chains
with repulsive monomer--monomer interactions behave
in qualitatively the same way as ideal chains.
The non-ideal chains must, however, be isolated, i.e., so
dilute even at the wall that the interaction between chains
is negligible, for this result to hold.
Interactions between chains changes the free energy of adsorption
and so the adsorption behaviour.

There is a thermodynamic transition only if the distribution $x(N)$ decays to
zero only at $N=\infty$. If there is some maximum chain length $N_c$
beyond which the distribution is cutoff, $x(N)=0$ for $N>N_c$,
then the transition is rounded off.
For $N_c\gg{\overline N}$, the normalisation of the distribution
is almost unaffected by the cutoff and
we can estimate $\rho_a$ by simply
replacing $\infty$ by $N_c$ as the upper limit of integration
in eq. (\ref{pgdef})
\begin{eqnarray}
\rho_a&\sim&
\frac{\left(\rho /{\overline N}\right)}
{{\overline N}^{-1}-(\epsilon/T-\epsilon/T_a)^2}
\left\{ \right.  1 - \nonumber\\
&&\left.\exp\left(
\left[(\epsilon/T-\epsilon/T_a)^2-{\overline N}^{-1}
\right]N_c\right)\right\}~~~~~T<T_a,\nonumber\\
\label{rhoafin}
\end{eqnarray} 
which can also be written as
\begin{eqnarray}
\frac{\rho_a}{\rho}&\sim&
\frac{1}
{1-{\overline N}(\epsilon/T-\epsilon/T_a)^2}
\left\{ \right.  1 - \nonumber\\
&&\left.\exp\left(
\left[{\overline N}(\epsilon/T-\epsilon/T_a)^2-1
\right]\frac{N_c}{{\overline N}}\right)\right\}~~~~~T<T_a.\nonumber\\
\end{eqnarray} 
The adsorption only depends on two parameters,
${\overline N}(\epsilon/T-\epsilon/T_a)^2$ and
$N_c/{\overline N}$.

There is no transition but for not too small ${\overline N}$
and $\epsilon/T-\epsilon/T_a$,
we see that the adsorbed density $\rho_a$ increases as
$\sim N_c(\epsilon/T-\epsilon/T_a)^2$. This should be compared with the
corresponding expression for monodisperse chains, eq. (\ref{rhoamono}).
The increase in adsorbed density when the polymer is polydisperse
is as fast as for monodisperse polymer of length $N_c$, equal to the longest
chains present in the polydisperse polymer.
This is perhaps
surprising, {\it a priori} we might have thought that the adsorbed
density increases about as fast as monodisperse chains of length equal to
the average length ${\overline N}$.
In fact polydisperse chains have a much higher adsorbed density than
monodisperse chains with the same average length ${\overline N}$ length.
This is shown in fig. 1, where the slope of the adsorbed density
for polydisperse chains (dashed curve) as
a function of the attraction energy over the thermal energy is very similar
to that of monodisperse chains with the length $N_c$ (dot-dashed curve)
and much larger than that of monodisperse chains with the average
length ${\overline N}$ (solid curve).

Physically, what is happening is that
most of the adsorbed monomers are from the largest chains of the distribution.
That most of the adsorbed monomers are from the longest chains can be
seen if we recall that the contribution to the adsorbed 
density of chains $\rho_a$ from chains of
length $N$ is given by the integrand of eq. (\ref{pgdef}).
For a cutoff distribution below the $T_e$
of a non-cutoff distribution with the same ${\overline N}$,
the integrand is an increasing function of $N$.

To summarise,
polydisperse ideal polymer chains show three qualitatively different
adsorption behaviours depending on the large $N$ tail
of the distribution function $x(N)$: (A) $x(N)$ decays
more slowly than exponentially in the $N\rightarrow\infty$ limit,
then the mixture behaves like infinite monodisperse chains; (B)
$x(N)$ decays exponentially in which case the adsorption density
diverges continuously, and (C) either $x(N)$ decays more rapidly
than exponentially or is cutoff at some maximum chain length, in
which case the mixture behaves like finite monodisperse chains.

To conclude, we have been able to classify polydisperse polymer solutions
into three classes, depending on their adsorption behaviour. The case
of an exponentially decaying distribution of chain lengths is especially
interesting as it behaves in a way which is qualitatively different
from that of monodisperse chains.
Sharp distinctions between monodisperse and the different
polydisperse distributions can only be drawn when there are infinite
chains present. Only then is a phase transition possible whose presence
or absence can be used to assign unambiguously a distribution to one of the
classes. However, at least when the polydisperse distribution function
$x(N)$ decays exponentially, even when the distribution is truncated
beyond some finite value $N_c$ there is a remnant of the behaviour
of the untruncated distribution.
Near the temperature $T_e$ where the adsorbed
density of the untruncated distribution would diverge there is changeover
in behaviour: above $T_e$ the shortest chains contribute most to the
adsorbed density $\rho_a$, below it the longest chains contribute most.

%\newpage

%\newpage
\section*{Figure Caption}

\vspace{0.1in}

\noindent
Fig. 1. The adsorbed density $\rho_a$ at the wall is plotted as a function
of the attraction energy over the thermal energy minus
its value at the adsorption transition for infinitely long chains.
The solid and dot-dashed curves are for monodisperse chains of lengths $N=$
50 and 150, respectively. The dashed curve is for polydisperse chains
distributed according to an exponential distribution of sizes
with an average chain length ${\overline N}=50$ and maximum
chain length $N_c=150$. At small $\epsilon/T-\epsilon/T_a$ the
adsorbed density for the polydisperse polymer is less than that for
monodisperse polymer of length $N=50$.

\end{multicols}

\begin{center}
\epsfig{file=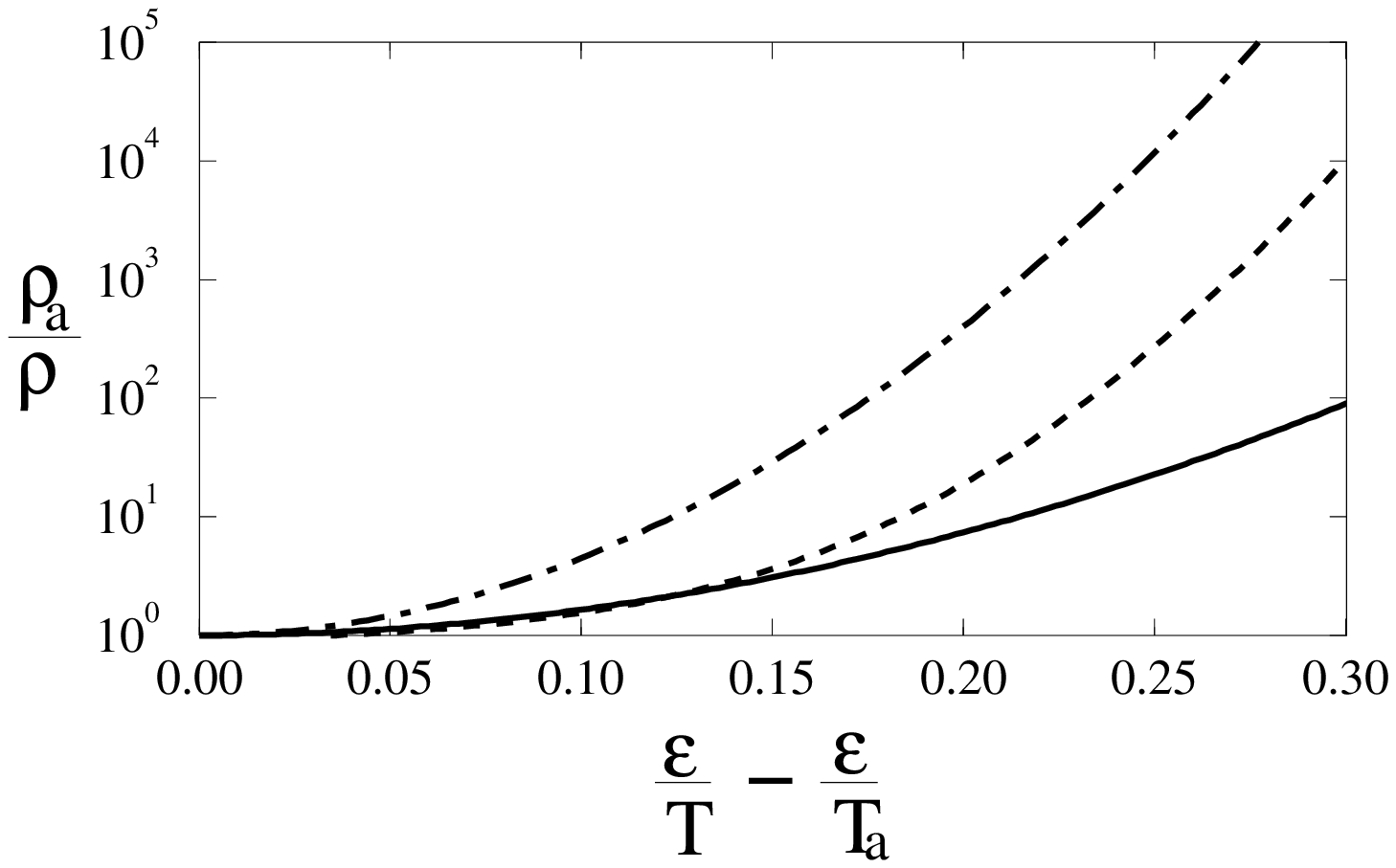,width=4.5in}
\end{center}


\begin{thebibliography}{99}

\bibitem{flory} FLORY P. J.,
{\it Principles of Polymer Chemistry}
(Cornell University Press, Ithaca) 1953.

\bibitem{degennes} DE GENNES P.-G.,
{\it Scaling Concepts in Polymer Physics}
(Cornell University Press, Ithaca) 1979.

\bibitem{rubin65} RUBIN R.,
{\it J. Chem. Phys.}, {\bf 43} (1965) 2392.

\bibitem{eisenriegler82} EISENRIEGLER E., KREMER K. and BINDER K.,
{\it J. Chem. Phys.}, {\bf 77} (1982) 6296.

\bibitem{eisenriegler83} EISENRIEGLER E.,
{\it J. Chem. Phys.}, {\bf 79} (1983) 1052.

\bibitem{eisenriegler} EISENRIEGLER E.,
{\it Polymers near Surfaces}
(World Scientific Press, Singapore) 1993.

\bibitem{roefs94} ROEFS S. P. F. M., SCHEUTJENS J. M. H. M. and
LEERMAKERS F. A. M.,
{\it Macromolecules}, {\bf 27} (1994) 4810.

\bibitem{solc70} \v{S}OLC K.,
{\it Macromolecules}, {\bf 3} (1970) 665.

\bibitem{sear} SEAR R. P.,
cond-mat/9811150 (http://xxx.lanl.gov).

\bibitem{salacuse82} SALACUSE J. J. and STELL G.,
{\it J. Chem. Phys.} {\bf 77} (1982) 3714.


\end{thebibliography}
\end{document}